\documentclass[%
pre, %
 amsmath,amssymb,
reprint,%
footinbib
]{revtex4-1}

\usepackage{amsmath}
\usepackage{amssymb}
\usepackage{amsfonts}
\usepackage{graphicx}
\usepackage{color}
\usepackage{hyperref}
\usepackage{bm}
\usepackage[english]{babel}
\usepackage{siunitx}
\usepackage[version=4]{mhchem}
\usepackage[utf8]{inputenc}
\usepackage[T1]{fontenc}
\usepackage{mathptmx}
\usepackage{etoolbox}

\renewcommand{\vec}[1]{\ensuremath{\mathbf{#1}}}

\newcommand{\kB}{\ensuremath{k_{\rm B}}}
\newcommand{\kb}{\ensuremath{\kB}}
\newcommand{\kT}{\ensuremath{\kb T}}

\newcommand{\broadwidth}{.45\textwidth}

\newcommand{\rgsq}{\ensuremath{R_{\rm g}^2}}

\newcommand{\thetitle}{Free Energy Landscape and Isomerization Rates of $\ce{Au4}$ Clusters at Finite Temperature}

\makeatletter
\def\@email#1#2{%
 \endgroup
 \patchcmd{\titleblock@produce}
  {\frontmatter@RRAPformat}
  {\frontmatter@RRAPformat{\produce@RRAP{*#1\href{mailto:#2}{#2}}}\frontmatter@RRAPformat}
  {}{}
}%
\makeatother

\begin{document}

\title{\thetitle}

\author{Jiale Shi}
\affiliation{%
  Department of Chemical and Biomolecular Engineering, %
  University of Notre Dame, %
  Notre Dame, IN 46556, USA
}

\author{Shanghui Huang}
\affiliation{%
    Department of Chemistry and Biochemistry, %
    University of Notre Dame, %
    Notre Dame, IN 46556, USA
}

\author{Fran\c{c}ois Gygi}
\affiliation{%
  Department of Computer Science, %
  University of California Davis, %
  Davis, CA 95616, USA
}

\author{Jonathan K. Whitmer}
\email{Author to whom correspondence should be addressed: jwhitme1@nd.edu}
\affiliation{%
  Department of Chemical and Biomolecular Engineering, %
  University of Notre Dame, %
  Notre Dame, IN 46556, USA
}
\affiliation{%
    Department of Chemistry and Biochemistry, %
    University of Notre Dame, %
    Notre Dame, IN 46556, USA
}

\date{\today}% It is always \today, today,

\begin{abstract}
In metallic nanoparticles, the cluster geometric structures control the particle's electronic band structure, polarizability, and catalytic properties. Analyzing the structural properties is a complex problem; the structure of an assembled cluster changes from moment to moment due to thermal fluctuations.  Conventional structural analyses based on spectroscopy or diffraction cannot determine the instantaneous structure exactly and can merely provide an averaged structure. Molecular simulations offer an opportunity to examine the assembly and evolution of metallic clusters, as the preferred assemblies and conformations can easily be visualized and explored. Here, we utilize the adaptive biasing force algorithm applied to first principles molecular dynamics to demonstrate exploration of a relatively simple system which permits comprehensive study of the small metal cluster $\ce{Au4}$ in both neutral and charged configurations. Our simulation work offers a quantitative understanding of these clusters' dynamic structure, which is significant for single-site catalytic reactions on metal clusters and provides a starting point for a detailed quantitative understanding of more complex pure metal and alloy clusters' dynamic properties.
\end{abstract}

\maketitle

\section*{Introduction}
In the assembly of nanoscopic and mesoscopic materials, clusters are important precursors where elemental units begin to aggregate.~\cite{lisimulation2013,bianchiselfassembly2015,meng2010free,huang2020surveying} 
The properties of initial aggregates that are formed control pathways available for subsequent assembly,~\cite{he2020colloidal,wang2017colloidal,chen2011supracolloidal} and therefore the ultimate structure of the assembled materials.
In the case of metallic nanoparticles, cluster geometry can impact the particle's electronic band structure, polarizability, and catalytic activity,~\cite{zhai2005unique,zhai2008chemisorption,liu2018optimum, austin2015au13,blavsko2017comparative,chen2008influence,lushchikova2019structures,blaskocomparative2017}  enabling many  applications in gas sensing, pollution reduction, biology, nanotechnology, and catalysis.~\cite{hakkinen2003structural,mills2003oxygen,daniel2004gold,Yoon2005charging,pyykko2008theoretical,hakkinen2008atomic,freund2011co,beret2014reaction,liu2017co,lang2017selective}
Metal clusters' geometric packing properties,~\cite{zhai2005unique} are very different from those of analogous clusters of isotropically attractive spheres.~\cite{huang2020surveying,meng2010free}
This arises from the directional interactions imparted by electron orbitals, which are predominantly anisotropic in nature.
For example, a four-particle colloidal cluster's stable structure is tetrahedral~\cite{meng2010free} while the equivalent stable structures of metals have different structures governed by electron-sharing geometries;~\cite{Imaoka2019Isomerizations,bonavcic2002density,boyukata2008structural} $\ce{Au4}$ tends to adopt either a two-dimensional rhombus ($\diamondsuit$) or a Y-shaped (Y) geometry.~\cite{bonavcic2002density} 
As the number of atoms in a metal cluster increases, the structures become more intricate and have more isomers.~\cite{goldsmith2019two}

Potential isomers of metal clusters may be identified by minimizing the energy in a DFT calculation without thermal fluctuations. Though these conformations can be long-lived, at finite temperature a typical metallic cluster is dynamic~\cite{liu2016mechanistic} and exhibits rapid transitions between different structural isomers.~\cite{Imaoka2019Isomerizations,wang2015dynamic,xing2006structural,Bulusu2006evidence,kryachoko2007magic,olson2007isomers}
Experimental work on small clusters with spatiotemporal microscopic resolution by Imaoka et al,~\cite{Imaoka2019Isomerizations} has revealed that the metallic cluster, $\ce{Pt4}$,  randomly walks through several isomers  and that the isomerizaton reactions follow simple first-order kinetics.
Computational studies have utilized classical molecular dynamics simulation~\cite{arslan1998dynamical,boyukata2008structural} and first-principles molecular dynamics simulation~\cite{goldsmith2019two, garzon1997ab, liu2000ab,zhang2019static, bravo1999non} to study the free energy profiles of neutral gas phase noble metal clusters and highlight the impact of temperature on the cluster stability. 
Most studies to date have focused on stable structures, and ignored kinetic concerns which are known to be vital for practical applications.~\cite{lee2021neural}
Recent simulation work by Wang et al,~\cite{wang2015dynamic} has, for example, demonstrated that $\ce{Au20}$ rearranges dynamically to form single-atom catalytic active sites. Such behavior is essential to the function of nanoparticle-based heterogenous catalysts. Understanding the kinetic behavior of clusters is thus key in determining the function of catalysts and engineering new materials. 

When acting as catalysts, metallic clusters frequently exchange electrons with a substrate or ligand during intermediate stages of a reaction.~\cite{camellone2009reaction} Therefore, the charge of the cluster is often dynamically altering during a catalytic process, and changes in different steps.~\cite{camellone2009reaction} 
Computational studies at 0 K  using DFT calculations which have been compared to infrared spectroscopy on experiments show that a charged cluster has a different ground state geometry than a neutral cluster.~\cite{zhai2008chemisorption,lushchikova2019structures} Thus, it is significant to study how the charge affects the free energy landscapes of metallic clusters at finite temperatures. Importantly, many of the properties of isomerization transitions, and how conformational states are influenced by the net charge on a metallic cluster, can be observed in small metallic clusters where they may be comprehensively understood.

\begin{figure}[h]
	\begin{center}
	\includegraphics[width=\broadwidth]{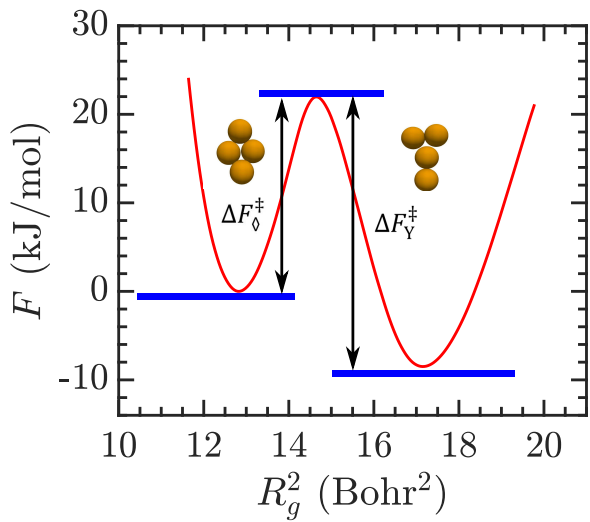}
	\end{center}
	\caption{Free energy landscape for \ce{Au4} at \SI{500}{\kelvin}, illustrating the relative height from each basin to the transition state ($\Delta F_i^{\ddagger}$). The Rhombus ($\diamondsuit$) and Y-shaped (Y) configurations of the \ce{Au4} cluster as observed in simulations are overlaid on the basins.}
	\label{fig:FEStemplate}
\end{figure}

Among all the noble metals, gold clusters are the most widely explored.~\cite{goldsmith2019two,Li2003Au} As the number of atoms in a gold cluster increases, the number of isomers increases significantly, resulting in more complex free energy landscapes.~\cite{goldsmith2019two} In this work, we choose $\ce{Au4}$ as a representative cluster and investigate the free energy landscape of gas phase neutral and charged $\ce{Au4}$ clusters to understand their thermodynamic and structural properties at different temperatures. The $\ce{Au4}$ cluster discussed previously is the smallest gold cluster which has multiple stable isomers, which are illustrated in Fig.\ref{fig:FEStemplate}. First-principles molecular dynamics (FPMD) simulations coupled with advanced sampling methods\cite{sevgen2018hierarchical} are applied to explore and analyze this system's stability and its isomerization rates. We additionally investigate the differences in thermodynamic and structural properties between neutral and charged metal clusters to look into the mechanisms by which the charge would affect the metal clusters' stability at finite temperature.

\section*{Methods}

The methods used in our calculation and information necessary to repeat or extend these calculations are discussed below.\footnote{All required input files are available via GitHub at \url{https://github.com/shijiale0609/clustersFES}} Biased first-principles molecular dynamics simulations (FPMD) were carried out using the Qbox code~\cite{gygi2008architecture} coupled with the SSAGES suite~\cite{sidky2018ssages} in client-server mode~\cite{sevgen2018hierarchical,govoni2021code,lee2021neural}. Electronic structure computations necessary to obtain the forces on nuclei that are evolved in MD are carried using density functional theory (DFT) with the PBE exchange-correlation functional. SG15 optimized norm-conserving Vanderbilt (ONCV) pseudopotentials~\cite{Martin2015Optimization} were used as implemented in the Qbox code.~\cite{gygi2008architecture} A plane-wave basis set expanded to an energy cutoff of 45 Ry\footnote{While the choice of the energy cutoff  has a nominal effect on ground state calculations for these two configurations (which are within approximately \SI{1}{\kJ\per\mol} of each other), we have verified that this choice does not significantly affect the free energy landscapes which are dominated more by entropy and multiplicity of locally available states.} represents the valence electrons, and the calculations include eight empty electronic states and Fermi smearing with an electronic temperature of 500 K. All calculations are performed without restricting spin. We do not explore effects of spin polarization in these calculations, though during preliminary work we compared the forces arising when considering spin polarization explicitly and have found its influence to be negligible, similar to the results of Liu, et al.\cite{liu2000ab} 

All FPMD calculations involve a periodic cubic cell with an edge length of  $a = \SI{13.92}{\angstrom}$, which is large enough that the $\ce{Au4}$ cluster does not interact with its own periodic images. A Bussi-Donadio-Parrinello (BDP) thermostat\cite{bussi2007canonical} is used to sample the canonical ensemble. As \ce{Au} is a relatively heavy atom, the MD simulation timestep is set to be \SI{1.935}{fs} for a balance of accuracy and efficiency. For each timestep, 10 self-consistent field calculations are performed, each involving five wave-function optimizations. Simulation settings for both charged and neutral clusters are identical, aside from modifying the overall net charge in Qbox to $q_{\rm tot}\in\left\{-1,0,1\right\}$. 

Elucidating isomerization transition rates requires efficient sampling of rare events, since free energy barriers must be crossed. Enhanced sampling calculations proceed by applying a bias to collective variables to speed up the exploration of the simulated systems. Collective variables (CVs), closely related to the concept of reaction coordinates, are a low-dimensional projection of the high-dimensional space of MD simulations, which can clearly distinguish reactants from products and quantify dynamical progress along the pathway from reactants to products.~\cite{peters2017reaction} Generally, this defines a vector valued function from the space of nuclear positions to the reduced CV space, $\vec{\xi}: \mathbb{R}^{3N} \rightarrow \mathbb{R}^d$, where $N$ is the number of atoms and $d$ the desired reduced dimensionality. For delineating two basins, it is typically sufficient to define a single collective variable (though this can lead to dynamical bottlenecks if poorly chosen\cite{Bussi2020}). Here, we opt for a single CV, the squared radius of gyration of the cluster $\rgsq$, defined as
\begin{equation}
    \rgsq = \frac{1}{4}\sum_{i=1}^{4}\left(\vec{x}_i-\vec{x}_{\rm CM}\right)^{2}\;.
    \label{eq:Rg2}
\end{equation}
\noindent $\rgsq$ captures the compactness of the cluster, and is thus sufficient to distinguish the two stable conformers. 
Here $\vec{x}_i$ and $\vec{x}_{\rm CM}$ refer to the position of each particle and the center of mass, respectively. This CV is able to distinguish the stable isomers of 
$\rm Au_{4}$ as shown in Fig.~\ref{fig:FEStemplate}, and has been demonstrated a useful variable in past investigations for driving transitions between compact and extended clusters.\cite{huang2020surveying}. Values at each basin in Fig.~\ref{fig:FEStemplate} are approximately $R^2_{g,\diamondsuit}\approx 12.8\ \rm(Bohr)^2$ and $R^2_{g,\rm{Y}} \approx 17.2\ \rm(Bohr)^2$. These positions necessarily shift slightly when the temperature or charge state are modified. 

Advanced sampling is carried out using the adaptive bias force (ABF) algorithm\cite{Darve2008} as implemented in SSAGES.\cite{sidky2018ssages}  The ABF method is constructed to swiftly surmount large barriers in free energy by negating the forces which act to restore a configuration to a stable basin; as such it is not only capable of accurately obtaining information about the relative free energies of basins, but can also accurately sample the barriers between them, enabling estimation of the transition rates using methods such as transition state theory (TST). Other sampling techniques could potentially be used. Prior works have applied replica-exchange Monte Carlo and molecular dynamics~\cite{beret2011free,goldsmith2019two} to study the free energy landscape of neutral equilibrium gold clusters ($\ce{Au5}$ - $\ce{Au13}$). These calculations require long equilibration times when the free-energy barriers are high between stable and metastable states.~\cite{machta2009strengths} Replica-exchange methods allow for an unbiased search of the potential-energy substrates from which gold cluster structures can be identified. From results by Goldsmith et al~\cite{goldsmith2019two} on the free energy landscape of $\ce{Au5}$, free energy barriers are not well-sampled especially when the barrier is very high, causing rate calculations to be inaccurate. Provided the CV used is a sufficiently good approximation of the true reaction coordinate, ABF can be used to swiftly scale and accurately determine free energy barriers in reacting processes; this is in particular beneficial when using FPMD as the underlying algorithm to sample configurations.

We utilize four walkers\cite{Raiteri2006} to accumulate the ABF bias on $\rgsq$, which is accumulated within the interval [11.5, 20.5] $\rm (Bohr)^2$ encompassing both basins of interest. When walkers wander outside this grid, a restoring force is applied outside of an initial unbiased buffer zone having a thickness of 0.3 $\rm (Bohr)^2$ to nudge the system back toward the region of interest. The one-dimensional region utilizes 90 bins, and the restoring forces are generated by a half-harmonic potential with a spring constant of \SI{0.1}{hartree/(Bohr)^4}, centered at the edges of the buffer zones. The ABF algorithm requires a minimum number $n_{\rm visit}$ of visits to each grid region before the full force is applied to avoid overemphasizing relatively rare states with large mean forces that can render a simulation unstable; here we set $n_{\rm visit} = 100$. Prior to applying the ABF algorithm, each walker is run for a short time ($\approx 1700$ MD steps) to obtain a locally equilibrated state. Enhanced sampling calculations are then performed for another $1 \times 10^6$ MD steps. We find that these settings are sufficient to explore this system's basins and transition state. 

With the converged free energy landscape, we are able to explore transition rates using transition state theory (TST). Specifically, here, we are interested in the isomerization rate of the \ce{Au4} cluster, but our discussion below is general to the computation of any reaction rate where the free energy landscape is known along a single collective variable or reaction coordinate. TST requires only minimal details about the  energy surface in the immediate vicinity of transition states barrier, reactant and product basins. The general expression for a reaction rate in TST is\cite{peters2017reaction}
\begin{equation}
    k_{TST} = \frac{\kT}{h} V_{0}^{n-1} \exp{(-\beta \Delta F^{\ddagger})}\;,
    \label{eq:gtst}
\end{equation}
\noindent where the volume per molecule $V_{0}$ corresponds to the reference concentration and to the volume in the translational partition functions, $n$ corresponds to the reaction order, and $\Delta F^{\ddagger}$ is the activation free energy. Eq.\ref{eq:gtst} tells us that to calculate the reaction rates for the first order reaction, one merely needs to know the free energy barrier of the transition states relative to the reactant basins. The information about free energy barriers is accessible using advanced sampling capable of driving successfully through these regions. These calculations are greatly facilitated using accurate free energy landscapes obtained through advanced sampling methods. It should be noted that more complex, multivariate expressions for TST can be used if the reaction coordinate is not accurately captured using a single CV; these could prove useful for studying isomerization rates between molecules in more complex clusters.

\section*{Results and Discussion}

\begin{figure}[h]
	\begin{center}
	\includegraphics[width=\broadwidth]{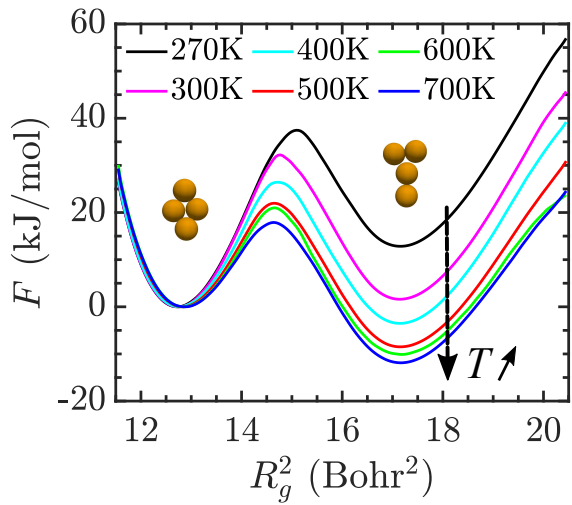}
	\end{center}
	\caption{Free energy landscapes of the $\ce{Au4}$ cluster with varying temperature. At lower temperatures, the rhombus isomer is the dominant state; this transitions to the Y-shaped isomer at $\approx \SI{400}{\kelvin}$. The free energy barriers also decrease with increasing temperature, and are easily activated as they comprise only a few $k_{\rm B} T$ at high temperatures, meaning the two cluster forms will readily interconvert.
	}
	\label{fig:Au4FES}
\end{figure}

Fig.~\ref{fig:Au4FES} shows a collection of converged free energy landscapes of neutral  $\ce{Au4}$ clusters at different temperatures between \SI{270}{K} and \SI{700}{K}. As can be reasonably hypothesized, the more compact, energetically favorable rhombus structure is favored at low temperatures, while the increased entropy within the Y-shaped structure is preferred as temperature is increased. It is clear, however, that both configurations are at least metastable at all temperatures explored here. The wells for each are relatively deep, with the free energy difference between the stable basin and transition state approximately \SI{20}{\kJ\per\mol} or more for each temperature studied. At the lowest temperature, $\Delta F^\ddagger_\diamondsuit \approx \SI{40}{\kJ\per\mol}$ and $\Delta F^\ddagger_{\rm Y} \approx \SI{25}{\kJ\per\mol}$, corresponding to $\approx 18 \kT$ and $\approx 11 \kT$, respectively. This marks each configuration as extremely stable on short timescales typical of molecular simulations, with sampling of these transitions unlikely to be accessible without ABF or similar algorithms. The absolute depths of the free energy wells remain on this order as the temperature rises, meaning the activation barrier relative to $\kT$ is reduced significantly at higher temperatures; $\Delta F^\ddagger_\diamondsuit$ becomes $\approx \SI{20}{\kJ\per\mol}$ at \SI{700}{\kelvin} ($\approx 3.5 \kT$), so that transitions between the two basins become more dynamically favorable. 

As temperature is increased, the relative stability of the rhombus decreases, with concomitant increase in the stability of the Y-shaped cluster. At $\approx \SI{400}{\kelvin}$, the Y-shaped isomer becomes the dominant configuration, due to its relatively higher entropy. Such effects have been previously noted to drive two-to-three dimensional transitions in $\ce{Au5}$ - $\ce{Au13}$ clusters~\cite{goldsmith2019two} as temperature is increased. We can analyse the temperature dependence of cluster stability by computing the equilibrium constant for the isomerization reaction. To obtain this quantity, we subdivide the collective variable domain into two regions belonging to the rhombus or Y-shaped clusters. Note that the equilibrium probability of finding the system in one state is calculated by integrating all the probabilities in that basin,~\cite{peters2017reaction}
\begin{equation}
    \tilde{P}(i) = \int_{\Xi_i} d\xi  \quad e^{-\beta F(\xi)}
    \label{eq:probability}\;.
\end{equation}
\noindent where $\xi \in \Xi_i$ defines the CV of interest, and $\Xi_i$ the fraction of the domain corresponding to basin $i$. Note that this probability is unnormalized; one must divide out by the full partition function to obtain this quantity. Importantly, the probabilities phrased in this way define the occupancy rate for a system to be in one of the stable basins. We separate the free energy landscape using the transition state and integrate each isomer's probability in the corresponding basin.  Therefore, we obtain an expression for the occupancy rate of each isomer as
\begin{equation}
\begin{aligned}
    r_\diamondsuit & = \frac{\tilde{P}_\diamondsuit}{\tilde{P}_\diamondsuit+\tilde{P}_{\rm Y}} 
    \quad \textrm{and} \quad 
    r_{\rm Y} & = \frac{\tilde{P}_{\rm Y}}{\tilde{P}_\diamondsuit+\tilde{P}_{\rm Y}} \;.
\label{eq:ration}
\end{aligned}
\end{equation}

\begin{figure}[h]
	\begin{center}
	\includegraphics[width=\broadwidth]{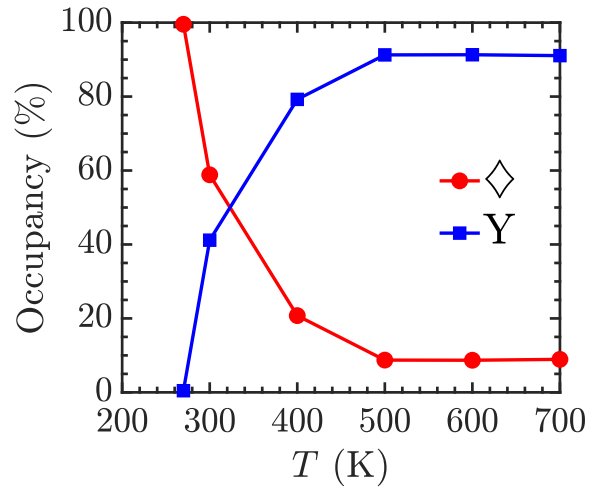}
	\end{center}
	\caption{Occupancy rates (\%) of $\diamondsuit$ (red circles) and Y (blue squares) isomers at different temperatures. At low temperature (\SI{270}{\kelvin}), $r_\diamondsuit \approx 100\%$, while with increasing temperature, the occupancy rate of the Y isomer increases essentially monotonically. At the highest temperatures, the occupancy rates of the two states stabilize at approximate fractions: $r_\diamondsuit \approx 9 \%$ and $r_{\rm Y} \approx 91 \%$.
	}
	\label{fig:occupationratio_Au4}
\end{figure}

From Fig.\ref{fig:occupationratio_Au4}, the rhombus isomer is clearly dominant at low temperature. At 270 K, the occupancy rate of the rhombus isomer is almost 100\%, while the occupancy rate of the Y-shaped isomer is almost zero. With increasing temperature, the occupancy rate of the rhombus isomer decreases, and the occupancy rate of the Y-shaped isomer increases. While the basins at \SI{400}{\kelvin} are approximately equally deep (see Fig.~\ref{fig:Au4FES}), there is a clear preference for the system to be in the Y basin due to it's wider character and greater resulting probabilistic weight. Interpolating linearly on our measurements predicts the isomers should be equally observed at approximately \SI{330}{\kelvin}. As the temperatures increase to \SI{500}{\kelvin} and beyond, the occupancy rates stabilize at approximate fractions $r_\diamondsuit \approx 9 \%$ and $r_{\rm Y} \approx 91 \%$. As temperature is increased further, the two basins would likely merge into a fluid-like single basin, while these would also compete with a dissociated gas-like state, though these transitions cannot be predicted from the data at hand.

\begin{figure}[h]
	\begin{center}
	\includegraphics[width=\broadwidth]{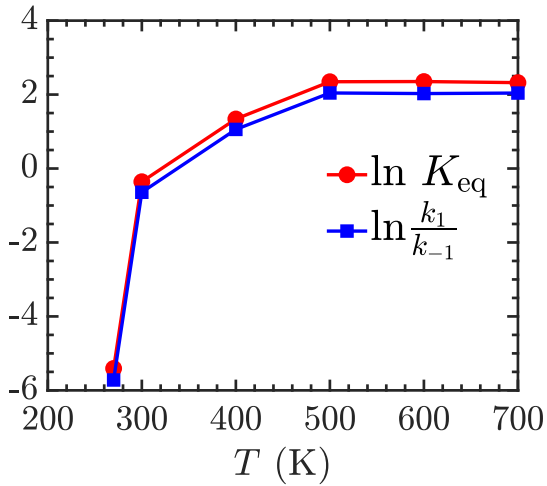}
	\end{center}
	\caption{The equilibrium constant in log-scale $\ln K_{\rm eq}$ (red circles) and 
	$\ln \frac{k_{1}}{k_{-1}}$ (blue squares) at different temperatures. These are numerically consistent, though $\ln \frac{k_{1}}{k_{-1}}$ is slightly lower, owing to the different accounting for entropy in the two calculations. See text for more details.
	}
	\label{fig:Kandkstst}
\end{figure}

The dynamic transition between two $\ce{Au4}$ isomers may be presented by a mono-molecular reaction with first-order kinetics. Previous experimental results have shown that isomerization with one small metal cluster with no degenerate states obeys a first-order rate law.~\cite{Imaoka2019Isomerizations} Therefore, we calculate the reaction equilibrium constant $K_{\rm eq}$ following the first-order reaction rule:
\begin{equation}
\begin{aligned}
    K_{\rm eq} &= \frac{r_\diamondsuit}{r_{\rm Y}}\;.
\label{eq:Keq}
\end{aligned}
\end{equation}
\noindent Here we have replaced the typical concentrations in a rate constant expression with the relative populations of each species, which are equivalent in a system of \ce{Au4} clusters, provided the system is not dense enough for clusters to interact. We plot the logarithm of the equilibrium constant$\ln K_{\rm eq}$ as a function of temperature in Fig.\ref{fig:Kandkstst}(red circles). With increasing temperature, $\ln K_{\rm eq}$ initially increases, and eventually becomes approximately flat, mirroring the flattening of the ratios in Fig.~\ref{fig:occupationratio_Au4}. Here we also observe that the transition between clusters should occur at $\approx \SI{300}{\kelvin}$ where $\ln K_{\rm eq}$ is zero.

\begin{figure}[h]
	\begin{center}
	\includegraphics[width=\broadwidth]{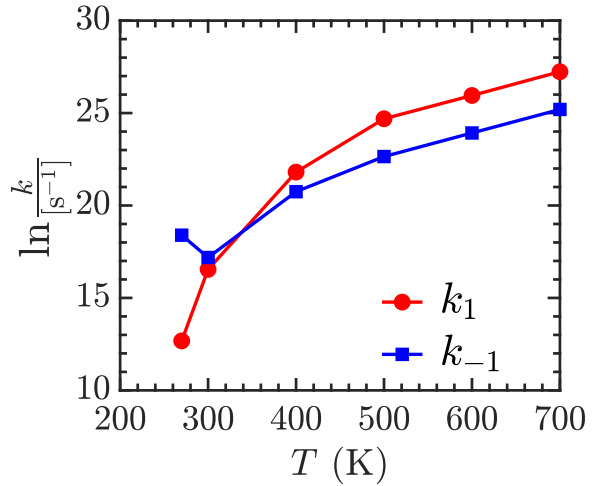}
	\end{center}
	\caption{The forward transition ($ \diamondsuit \rightarrow {\rm Y}$) reaction rate $k_{1}$ and the reverse transition (${\rm Y}  \rightarrow \diamondsuit$) reaction rate $k_{-1}$ at different temperatures. In general, with temperature increasing, $k_{1}$ and $k_{-1}$ both increase, as basins become less stable locally.
	}
	\label{fig:ktst_Au4}
\end{figure}

Transition state theory~\cite{peters2017reaction} may be used to calculate the transition rates at different temperatures.
The dynamic transition between the metal cluster isomers is a first-order reaction,~\cite{Imaoka2019Isomerizations} 
thus the expression of reaction rate $k$ is
\begin{equation}
    k = \frac{\kT}{h} \exp{(-\beta \Delta F^{\ddagger})}\;.
    \label{eq:tst}
\end{equation}
\noindent Phrasing the reaction as
\begin{equation}
    \diamondsuit \ce{<=>[\textit{k}_1][\textit{k}_{-1}]}
    {\rm Y}\;,
\end{equation}
\noindent the forward and reverse rates are plotted in Fig.\ref{fig:ktst_Au4}. 
In general, with temperature increasing, $k_{1}$ and $k_{-1}$ both increase, as is expected, since $\Delta F_i^\ddagger$ is relatively constant in magnitude over this temperature range, resulting in decrease of $\Delta F^\ddagger_i/kT$. The exception is the reverse reaction $k_{-1}$ at \SI{270}{\kelvin}, which is significantly faster than the same reaction at \SI{300}{\kelvin} due to the much lower free energy barrier for the reverse reaction at \SI{270}{\kelvin}. This type of feature would be expected for systems which are only marginally stable, and for which the second stable basin may disappear in this CV for low temperatures. This perspective is supported by the occupancy statistics, which show the Y isomer to be essentially unpopulated at \SI{270}{\kelvin}.

To close these calculations, we note that we can compare the results of transition state theory to our occupancy-based results for $K_{\rm eq}$ using the relationship
\begin{equation}
     \ln K_{\rm eq}=  \ln \frac{k_{1}}{k_{-1}}\;.
     \label{eq:Kandk}
\end{equation}
\noindent This quantity is plotted alongside our prior results in Fig.~\ref{fig:Kandkstst}. We note that the two are in broad agreement, but that the the equilibrium constant that we calculate from Equ.~\ref{eq:Kandk} is a bit smaller in comparison to that calculated from occupancy statistics. This is primarily due to the fact that $K_{\rm eq}$ from ABF takes into account the relative weight of the whole basin, which favor more the entropy found in the Y-shaped isomer basin, though there could also be some small effects due to misalignment of the true reaction coordinate with $\rgsq$ that would cause $\Delta F^\ddagger_i$ to differ from the true activation energy. Since the differences are relatively small here, we can conclude that $\rgsq$ is a reasonable reaction coordinate for this transition. 

To put these results within the context of typical ab-initio simulations, we utilize the widely applied harmonic approximation method (HA)~\cite{peters2019calculating,sethna2021statistical} to calculate the free energy difference $\Delta F = F_{\diamondsuit}- F_{\rm Y}$ to illustrate the differences which arise between the HA and more comprehensive sampling methods such as ABF. The HA assumes the conformational probability (partition function) may be broken down into contributions from translation $t$, rotation $r$ and vibration $v$ along with an overall weight from the ground state energy $U_{0,i}$ of a given conformation according to Peters~\cite{peters2019calculating} and Meng.~\cite{meng2010free}

\begin{equation}
    P_i \propto e^{-\beta F_i} = Z e^{-\beta U} = Z_{t} Z_{r} Z_{v} e^{-\beta U_{0,i}}
\end{equation}

\noindent where $i$ labels the conformation of interest. The translational components are essentially equal, but rotational, vibrational, and ground state energies are necessarily different because of differences in the underlying symmetries, inertial axes, and potential energy landscape of each conformation. For the rotational component,~\cite{cole2001automatic} we use the online ABC rotational constant calculator~\cite{abcrotation} to get the relative rotational constants A, B, C for $\diamondsuit$ and $\rm Y$ to plug into the relation

\begin{equation}
Z_{r} = \frac{\sqrt{\pi (kT)^3}}{\sigma \sqrt{ABC}}\;.
\end{equation}

\begin{figure}[h]
	\begin{center}
	\includegraphics[width=\broadwidth]{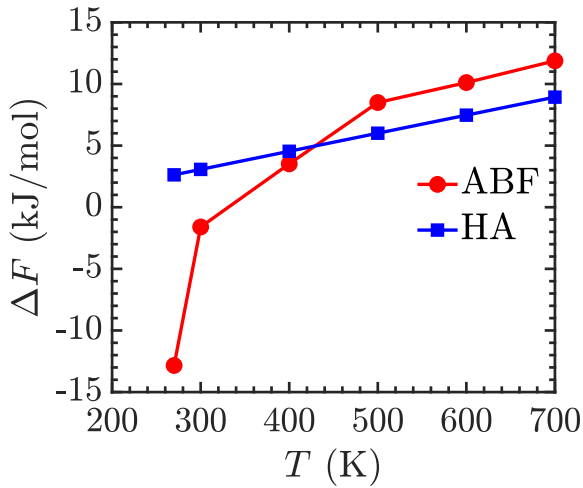}
	\end{center}
	\caption{$\Delta F = F_{\diamondsuit}- F_{\rm Y}$ at different temperatures, calculated by ABF sampling free energy method (red dots) and harmonic approximation (HA) method (blut squares) 
	}
	\label{fig:DF_compare}
\end{figure}

\noindent The vibrational component is calculated using Qbox to find the vibrational modes of the cluster, using finite difference calculations of the dynamical matrix, keeping the six highest frequency modes.\footnote{All other modes are accounted for in the translational and rotational partition functions of the cluster. The six vibrational frequencies for $\diamondsuit$ are 35, 69, 86, 86, 143, and 162 $\SI{}{\per\cm}$ and the six vibrational frequencies for $\rm Y$ are 21, 37, 86, 94, 156, and 203 $\SI{}{\per\cm }$} We then use the quantum mechanical expression\cite{peters2019calculating} for the vibrational free energy, 
\begin{equation}
    F_{\rm v}^{\rm QM} = \beta^{-1}\sum_{i}\ln \left(\frac{1-\exp{(-\beta h v_{i})}}{\exp{(-\frac{1}{2}\beta h v_{i})}}\right)\;,
\end{equation}

\noindent where the index $i$ runs over the different vibrational modes of interest. Finally, adding in the ground state energy difference computed using Qbox, $\Delta U_0 = U_{0,\diamondsuit}- U_{0,{\rm Y}}\approx  \SI{-1.2742} {\kJ\per\mol}$, we arrive at the result plotted in Figure~\ref{fig:DF_compare}. Interestingly, while the ABF sampling shows a clear transition between the two clusters, the harmonic approximation uniformly favors the ${\rm Y}$-shaped conformation, illustrating the key role played by entropy in these transitions, and the importance of using full free energy landscape sampling even in {\it ab initio} contexts.

% In Fig.\ref{fig:DF_compare}, we plot the $\Delta F$ vs $T$ with ABF sampling method and harmonic approximation (HA) method. We find that with HA $\Delta F$ follows a linear relationship with T. At high temperature, the slopes are almost the same and there is an almost constant difference.}  

\subsection*{Charged Clusters}

As noted, in real catalysts, metal clusters are not always neutral, and in many processes,~\cite{green2021atomic} the charge distribution will change dynamically due to the ligands and substrate interactions involved in the catalytic cycle.As an example, Qiao et al.~\cite{qiao2011single} find that in the $\ce{Pt/FeO_x}$ system, electrons transfer from a Pt atom to the $\ce{FeO_x}$ substrates, resulting in positively charged single Pt atoms $\ce{Pt+}$,  which exhibits remarkable catalytic performance for CO oxidation and  preferential oxidation. Similarly, Camellone et al~\cite{camellone2009reaction} have found that in catalytic oxidation of CO using $\ce{Au/CeO2}$, the active gold atoms within the catalyst are positive ions. These species activate molecular CO and catalyze its oxidation. Interestingly, oxygen vacancies can attract supported positively charged  $\ce{Au+}$ and turn them into negatively charged $\ce{Au^{-}}$.  Since charge shifting plays a prominent role in catalytic processes, it is vital to investigate the effect of charge on the free energy of metal clusters. Here, we explore the influence that charge has on the conformational properties of \ce{Au4} clusters in comparison to the neutral species.

\begin{figure}[h]
	\begin{center}
	\includegraphics[width=\broadwidth]{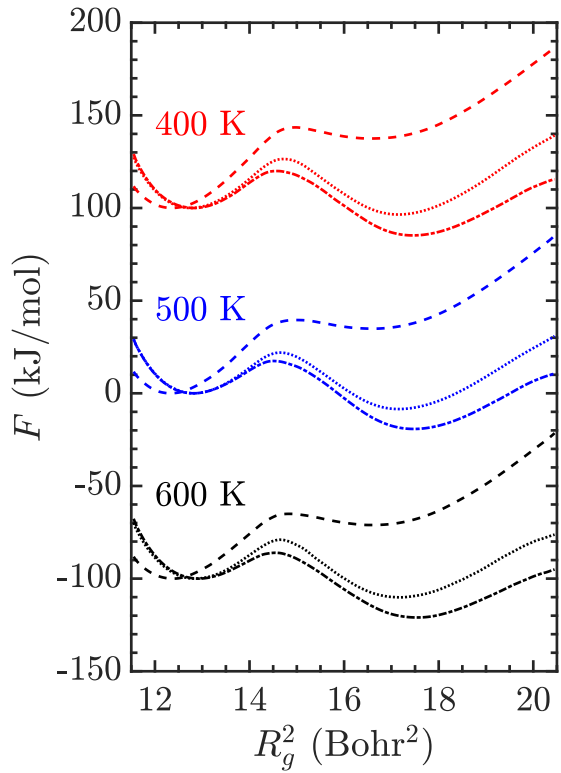}
	\end{center}
	\caption{Free energy landscapes of $\ce{Au4}$(dotted line, $\cdot \cdot$), $\ce{Au4-}$ (dash-dotted line, $-\cdot$), and $\ce{Au4+}$(dashed line, $--$) at 400 K (red), 500 K(blue), and 600 K(black). 
	}
	\label{fig:FES_charge}
\end{figure}

We summarize our results in Fig.~\ref{fig:FES_charge}. Some clear differences are manifest in the charged systems. We focus here on \ce{Au4+} and \ce{Au4-}, as previous researchers only find at most one positive charge or one negative charge on a small Au cluster for the catalysis process of Au clusters.~\cite{camellone2009reaction,Khetrapal2018Determination,forstel2019optical,jaeger2016photodissociation} The first observed difference is the $\rgsq$ values of the two isomers. When neutral $\ce{Au4}$ loses one electron and becomes positively charged, the $\rgsq$ values of the two isomers becomes smaller than those of the neutral $\ce{Au4}$ cluster. This is somewhat surprising, since one would expect charges centered at the nuclei to be more repulsive thus expanding the cluster. However, this type of reasoning is inherently classical; the remaining electrons in the positive cluster create a cloud which is effectively smaller than the neutral case, and these localized negative charges act to draw the nuclei closer together. This leads to the decreasing of the $\rgsq$ values of the whole cluster. Interestingly, when neutral $\ce{Au4}$ gains one extra electron and becomes negative charged, the $\rgsq$ values expand slightly, owing to the environment of the electron cloud pulling nuclei further apart. Notably, the effects are more pronounced for \ce{Au4+} than for \ce{Au4-}. It is not clear how much these effects will generalize to larger clusters and higher valence, though it is likely the effects are more pronounced in the case of \ce{Au4} than they would be in a larger cluster. Other factors, like the impact of electrostatic repulsion on the cluster size, must then also be considered. 

\begin{figure}[h]
	\begin{center}
	\includegraphics[width=\broadwidth]{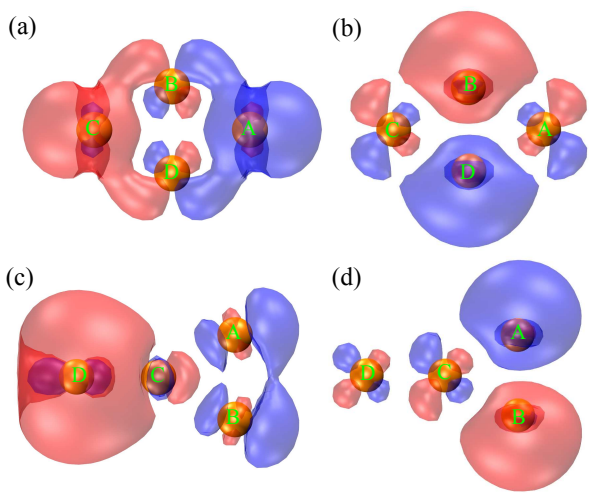}
	\end{center}
	\caption{Molecular orbitals of $\ce{Au4}$ at the ground state (0 K). Rhombus isomer's HOMO (a) and LUMO (b). Y-shaped isomer's HOMO (c) and LUMO (d).
	}
	\label{fig:orbital}
\end{figure}

The explanation may be bolstered by analysing the HOMO (Highest Occupied Molecular Orbital) and LUMO (Lowest Unoccupied Molecular Orbital) of the $\ce{Au4}$ clusters.~\cite{liu2013structure} For the rhombus isomer, the HOMO and LUMO have a similar anti-bonding "p-like" symmetry, but are located on different pairs of atoms, shown in Fig.\ref{fig:orbital} (a) and (b). The HOMO orbital is mostly localized on atoms A and C, whereas the LUMO is mostly localized on atoms B and D. Adding one extra electron to the LUMO increases the occupation of an anti-bonding state between B and D, which makes the bond length of B and D larger, resulting in an increased $\rgsq$. Removing one electron from the HOMO will reduce the occupation of an anti-bonding state between A and C, leading to the a smaller bond between A and C resulting in a smaller $\rgsq$. The Y-shaped isomer's  HOMO and LUMO are shown in Fig.\ref{fig:orbital} (c) and (d). The HOMO orbital is mostly localized on A, B and D, whereas the LUMO is mostly localized on A and B. Adding one extra electron to the LUMO increases the occupation of an anti-bonding state between A and B, makes the bond length of A and B larger resulting in a larger $\rgsq$.~\cite{liu2013structure} Removing one electron from the HOMO will reduce the occupation of an anti-bonding state between A and D (or B and D), makes the bond length of A and D (or B and D) smaller and resulting in a smaller $\rgsq$. This additionally breaks the symetry of the Y-shaped isomer.

A pronounced effect is also observed in the relative free energy of the two isomers. Compared with the neutral $\ce{Au4}$ cluster, the relative free energy of Y-shaped isomer in $\ce{Au4+}$ increases significantly, and the Y-shaped isomer is metastable (and only weakly stable) at all studied temperatures; this is commensurate with our explanation relating cluster size to the electron cloud around the metal nuclei, which is unable to stretch as far when the cluster is positively charged. Similarly, the effect of the negative charge is to favor less compact structures, destabilizing the rhombus isomers at all studied temperatures. The relative difference in depth can be quite significant, meaning a cluster that becomes charged when in a stable rhombus (or Y shaped) configuration will experience a significant thermodynamic driving force toward the other conformation when gaining (or losing) an electron. Such effects could have pronounced influence on the effectiveness of nanoparticles in heterogenous catalysis, or the relative structure of nanoparticles capped by specific ligands,~\cite{scalise2018surface,chevrier2018structure,salorinne2020synthesis,Rusishvili2020Stoichiometry,jiang2020Cucurbiturils} and signify an important avenue for subsequent research.

\section*{Conclusion}
In this work, we investigate the thermodynamic and structural properties of $\ce{Au4}$ clusters (neutral, monoanionic, monocationic) in the gas phase by surveying the free energy landscape and the isomerization rates. From this, a few general conclusions may be reached. One is that temperature is an important factor in determining the likely conformations, as it greatly affects the stability of $\ce{Au4}$ clusters with higher temperatures favoring larger entropy within the Y-shaped isomer. Exploring these free energy landscapes, we are able to quantitatively calculate the equilibrium constant $K_{\rm eq}$ and transition rates $k_{1}$ and $k_{-1}$ of the $\ce{Au4}$ isomerization reaction at different temperatures, observing an intriguing nonmonotonicity, as one of the clusters becomes less mechanically stable at lower temperatures. Further, a cluster's charge significantly affects its preferred equilibrium energy in surprising ways, concomitantly affecting the relative free energy and stability of the two isomers. The dynamic charge distribution change in catalysis processes can impact the metal cluster's geometric structures. These observations are enabled by exploration of the full free energy landscape available to the gold clusters using {\it ab initio} molecular dynamics and the ABF method; accounting for thermalization and entropy using only density functional theory and the harmonic approximation would have led to different conclusions.

Many other essential factors in the behavior of metal clusters remain to be explored in relation to experimental observations that typically take place near an interface, such as the influence of substrate geometry~\cite{camellone2009reaction} and the presence of defect points.~\cite{Imaoka2019Isomerizations} Apart from interconversion within one cluster's multiple isomers, catalysis processes have more complicated cluster-to-cluster transformations,~\cite{ren2018cluster} which make interesting targets for applications of these coupled {\it ab initio} and advanced sampling methods. Importantly, industrial catalysis widely uses alloy clusters,~\cite{li2021hydrogen,leon2018probing,Jian2017Nb} and handling the thermodynamic and structural properties of complex alloys is thus essential and challenging. New advances in artificial intelligence and machine learning,~\cite{lamoureux2020artificial,deringer2021origins} have offered new solutions to study structural and transitional properties on complicated metal and alloy systems with faster computational speed while retaining accuracy, and offer the potential to build on the framework established here to study complex physical processes in nanoparticles.

\section*{Code Availability}
Example scripts and information necessary to run the examples contained in this article are posted at \url{https://github.com/shijiale0609/clustersFES}. The Qbox code is available at \url{http://qboxcode.org}. SSAGES may be downloaded from \url{http://ssagesproject.github.io}.

\section*{Acknowledgement}
JS, FG, and JKW acknowledge the support of MICCoM, the Midwest Center for Computational Materials, as part of the Computational Materials Sciences Program funded by the U.S.  Department of Energy, Office of Science, Basic Energy Sciences, Materials Sciences  and Engineering Division, for the development of algorithms and codes used within this work. Contributions of SH and JKW were additionally supported by the United States National Science Foundation (Award No. DMR-1751988). JS and JKW acknowledge the use of computational resources at the Notre Dame Center for Research Computing (CRC). JS and JKW thank Dr. Elizabeth M. Y. Lee (U. Chicago) for assistance and discussions.

\bibliographystyle{apsrev}
\bibliography{ref}

\begin{thebibliography}{79}
\bibitem[{Noteii()}]{Noteii}
\bibinfo{note}{While the choice of the energy cutoff has a nominal
  effect on ground state calculations for these two configurations (which are
  within approximately \SI {1}{\kJ \per \mol } of each other), we have verified
  that this choice does not significantly affect the free energy landscapes which
  are dominated more by entropy and multiplicity of locally available states}.
\bibitem[{Noteiii()}]{Noteiii}
\bibinfo{note}{All other modes are accounted for in the
  translational and rotational partition functions of the cluster. The six
  vibrational frequencies for $\diamondsuit $ are 35, 69, 86, 86, 143, and 162
  $\SI {}{\per \cm }$ and the six vibrational frequencies for $\protect \rm Y$
  are 21, 37, 86, 94, 156, and 203 $\SI {}{\per \cm }$}.
\end{thebibliography}

\end{document}